\documentclass[manuscript,screen]{acmart}
\AtBeginDocument{%
  \providecommand\BibTeX{{%
    \normalfont B\kern-0.5em{\scshape i\kern-0.25em b}\kern-0.8em\TeX}}}

\setcopyright{acmcopyright}
\copyrightyear{2023}
\acmYear{2023}
\acmDOI{XXXXXXX.XXXXXXX}

\begin{document}

\title{Ball-AR: Fostering Playful Co-Located Interaction Through Environment-centric Physical Activity with AR}

\author{Arnav Kumar}
\affiliation{%
  \institution{Princeton University}
  \city{Princeton, NJ}
  \country{USA}}

\author{Andrés Monroy-Hernández}
\affiliation{%
  \institution{Princeton University}
  \city{Princeton, NJ}
  \country{USA}
}

\renewcommand{\shortauthors}{Kumar and Monroy-Hernández}

\begin{abstract}
  We present Ball-AR, an augmented reality (AR) game where two players in the same physical space attempt to hit each other with virtual dodgeballs overlaid on the physical world. Researchers have studied AR's potential for fostering co-located interaction and physical activity; however, they have not investigated the impacts of physical activity and physical environment on user experiences and interaction. We created an AR dodgeball game centered around encouraging physical activity and harnessing the physical environment. We then evaluated the game with five dyads to analyze the impacts of these design choices on the quality of gameplay and interaction between players. We found that physical activity and the shared physical space created memorable experiences and interactions among participants, although participants desired a more augmented and immersive experience. 
\end{abstract}



\begin{teaserfigure}
  \includegraphics[width=\textwidth]{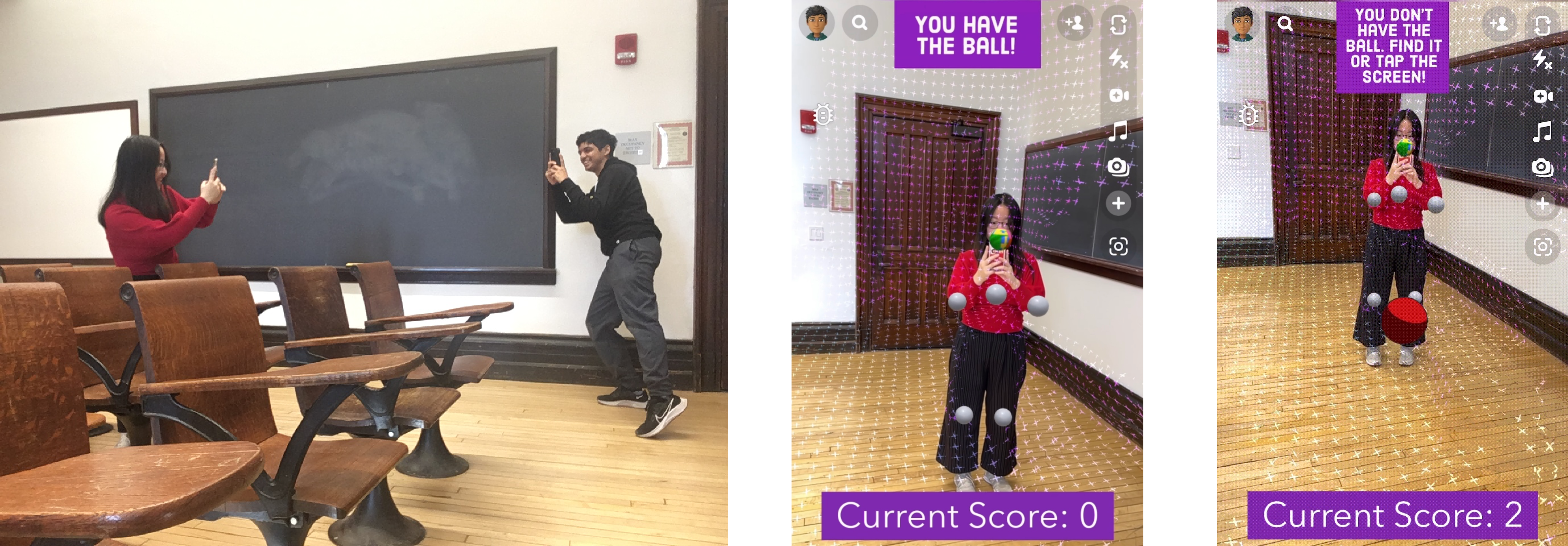}
  \caption{\textbf{The Ball-AR system in use. Ball-AR requires two players in the same physical space, and utilizes surface detection and body tracking for ball movement.}}
  \Description{The Ball-AR system in use. Ball-AR requires two players in the same physical space, and utilizes surface detection and body tracking for ball movement.}
  \label{fig:teaser}
\end{teaserfigure}


\maketitle

\section{Introduction}
In this paper, we discuss mobile AR game design by creating Ball-AR, an AR game inspired by dodgeball. Specifically, we study the impacts of game design choices based on physical and social areas that we believe AR can influence positively.

Our first motivation is to promote physical activity and encourage movement. As a result of the COVID-19 pandemic, physical activity levels have decreased an estimated 41\% \cite{wilke_pandemic_2021}. Conversely, in this same time frame, we saw a 75\% increase in gaming activity \cite{king_problematic_2020}. These behavior shifts motivated the idea to study how a game can encourage physical activity. In addition, AR games allow the ability to move while playing, so a mobile AR game provides a great opportunity to encourage physical activity. 

The second motivation is to foster co-located interaction. Recent research has shown that mobile devices can negatively impact communication and experiences between people in the same space \cite{przybylski_can_2013-1}. However, mobile AR games can be beneficial in fostering co-located interaction since they can facilitate communication and create fun, memorable experiences \cite{dagan_project_2022}. Another benefit AR can have on co-located interaction is its usage of the physical environment, as research has shown that the shared physical features of an environment can induce interaction between individuals by interrelating them through a shared space \cite{amran_effect_2020}.

We envision using AR to support physical activity in shared physical spaces. To do so, we created Ball-AR, an AR game that builds on and extends previously researched design choices. We foster movement by requiring players to move quickly, encourage social interaction by requiring sharing a physical location, and connect players physically by utilizing the physical features of the common environment.

We evaluated how well Ball-AR deals with fostering movement, encouraging social interaction, and utilizing the physical environment. We also study how the design choices affect participants' enjoyment of the game and their interaction with their opponents. 

We learned that the combination of a shared physical space and physical movement created enjoyable interactions and that competition, hidden information, and the physical elements of AR were effective in increasing communication and physical activity. However, participants did not feel connected to each other through the shared physical environment's features and desired more augmentation rather than utilization of just the physical environment.
\section{Related Work}
Recent research has shown that people are increasingly lonely and craving interaction \cite{weissbourd_loneliness_2021}. Mobile technology may add to this disconnect, as it can negatively affect communication and interactions in the same room \cite{przybylski_can_2013}. In addition, people have become increasingly sedentary \cite{wilke_pandemic_2021}. Unfortunately, mobile technology may also influence this, as studies have shown negative correlations between smartphone usage and physical activity levels\cite{school_of_physical_education_and_sports_van_yuzuncu_yil_university_van_turkey_effects_2020}.

With an estimated 5 billion people owning a mobile device, it becomes important to use these devices to encourage interaction and physical activity \cite{perrin_mobile_2021}. As a result, more and more mobile technologies have emerged to do so. Regarding fostering social interaction, video communication platforms like Zoom and Twitch have decreased loneliness-related outcomes, and improved social interactivity \cite{kaveladze_social_2022}. For physical activity, wearable technologies such as smartwatches or smartphones with mobile apps have positively impacted physical activity and weight control \cite{yen_effectiveness_2019}. However, not enough mobile technologies support both  co-located interaction and physical activity. 

To build our system, we focus on using AR, due to its known affordances of social interaction and physical activity. Previous works have shown that AR can foster co-located interaction due to its usage of shared spaces and embodied interaction. One previous work, Project IRL studied the influences of different design choices and features of mobile AR games on co-located interaction, some of which informed the design of Ball-AR \cite{dagan_project_2022}. Another work, BragFish, studied specifically the impacts of shared virtual spaces on the physical and social interactions among players \cite{xu_bragfish_2008}. Both studies noted the potential benefits of AR games in fostering co-located interaction. However, these works did not center on combining physical activity and co-location in AR.

In addition, other works have shown the potential for AR to encourage physical activity. The mixed reality game VRabl immersed players in a virtual space and explored the impacts of different design features on inducing physical activity \cite{buckers_vrabl_2018}. Game design choices from this study were used in Ball-AR. However, this study similarly did not research how inducing physical activity related to the quality of gameplay or interaction. 

As a result, we study game features that encourage co-located interaction and physical activity with the goal of providing insight for future mobile AR game design. To do so, we design a mobile AR game, Ball-AR, and analyze how AR's usage of the shared physical space and embodied interaction can benefit co-located experiences for players.

\section{Ball-AR System}

\begin{figure}[ht]
  \centering
  \includegraphics[width=\textwidth]{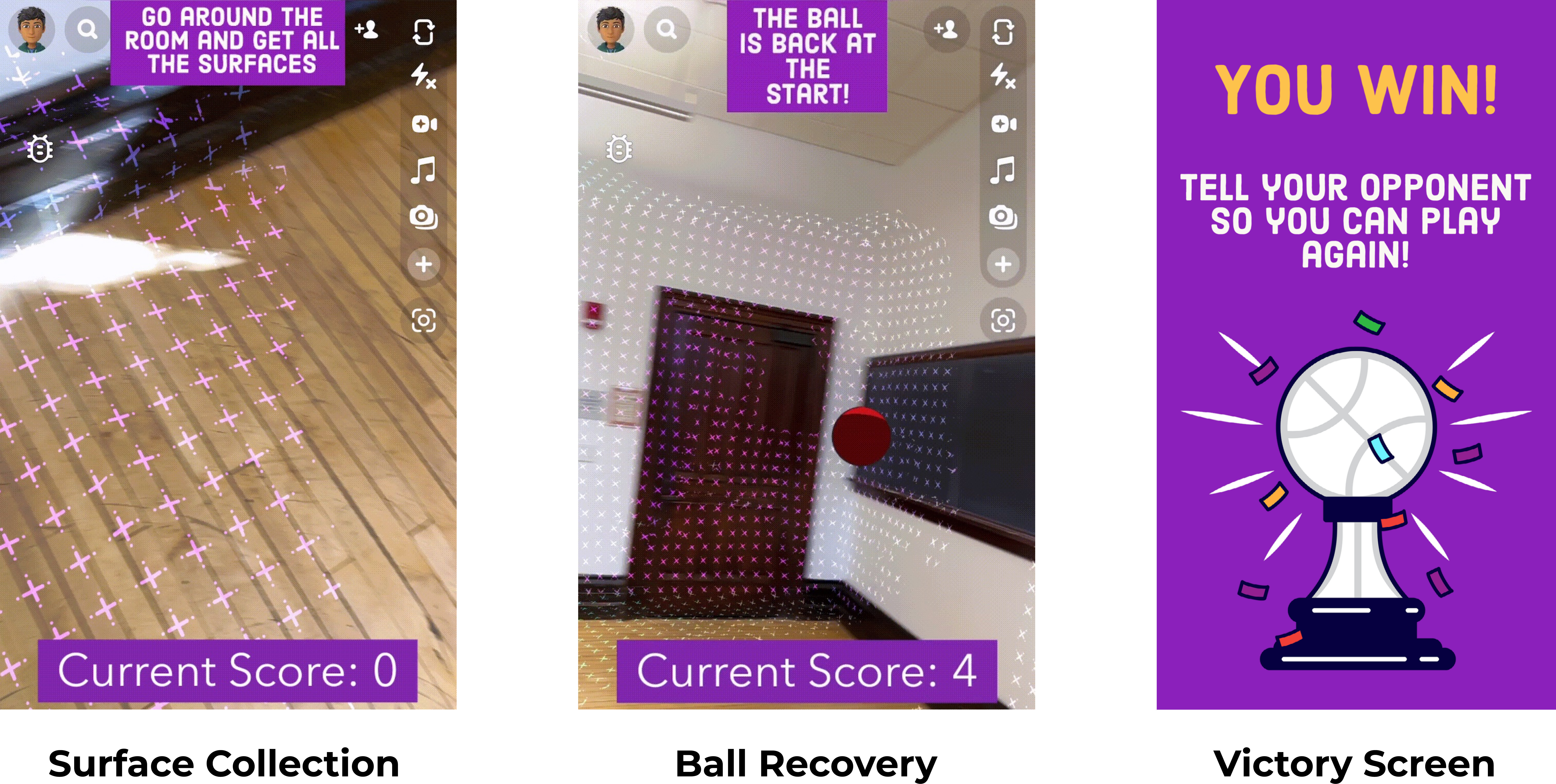}
  \caption{Description of different features within Ball-AR. Many features were designed to encourage movement or environment utilization.}
\end{figure}

Ball-AR is a multiplayer AR dodgeball game designed to be played by users in a shared physical space. To start the game, both players load the game on their mobile phones and stand in opposite corners of the room.

Players begin by first scanning the surfaces in the shared physical space. This entails walking around the room and pointing their mobile device at all physical surfaces, which are then captured by the lens. After collecting the environment, each user stands in the corner of the room, which is then deemed the starting spot. From this point on, each user now can throw their augmented ball at their opponent by tapping the screen. The ball can bounce off of any of the surfaces before hitting the opponent. If the ball hits three surfaces without hitting the opponent, the player must collect the ball to throw it again. If the ball hits the opponent without touching a surface, the player gets one point. In addition, the player gets an extra point for each surface hit before hitting the opponent (with a maximum of three surfaces). For example, if the player hits two walls and then their opponent, their score increases by three. The first player to reach ten points wins.

The game is designed to encourage physical activity and utilization of the shared physical environment to study impacts on interaction. For example, since each user only has one ball, they must physically move around the room in order to collect it. The aspect of competition adds a sense of urgency to the game, imploring users to rush to collect their ball for a better chance at victory. In addition, the extra points awarded based on surfaces hit before hitting an opponent are included to ground the users in the shared physical space and to focus on the physical features of the environment.

Ball-AR was designed with Lens Studio and is accessible via the Snapchat app. More specifically, the World Mesh and Body Tracking features within Lens Studio proved very useful in scanning a physical environment's surfaces and determining when the ball was hitting a player. 

\section{Evaluation}
We recruited ten university students through email to play Ball-AR in pairs. We then brought the participants to a predetermined location with ample surfaces and no major obstructions. Each participant played a best-of-three match of Ball-AR against another participant. Then, each participant was interviewed individually. We conducted semi-structured interviews (see protocol in Appendix) with participants and focused on concepts related to overall enjoyment, physical activity, environment utilization, and quality of interaction. The interview included questions such as "How did physical activity affect your overall experience?" and "Did you make use of the ability to bounce the ball off physical surfaces? What were your opinions on this feature?". 

\section{Study Results}
\subsection{Participants felt shared physical space and movement fostered greater interactions, though some desired a single-player experience as well}
Many participants felt that being in the same room as their opponent created a more fun experience. \textit{"It was cool to be in the same space and it made the game more fun to play by being in the presence of the person you're playing against"} (P6). In addition, the aspect of both players physically running around in the same room created enjoyable interactions between players. \textit{“The game was kind of a frenzy because I was just running around because I didn't want to get hit. [P7] started laughing, probably because I was so hard to actually hit”} (P8). However, many participants voiced a desire to have a single-player option as well to be able to play even when alone. \textit{"Having a game with both single-player and multiplayer would be nice. Multiplayer caters more to playing with friends but single-player could make more usable as a mobile game and for exercise by myself"} (P3). While participants enjoyed the physical interactions with their opponents and the shared physical space, many still viewed mobile games as a single-player experience.

\subsection{Hidden information increased communication and activity for participants}
Within Ball-AR, players could not see each other's scores or see the opponent's ball, although they were in the same room. As a result, participants communicated throughout the game, shouting scores and exclaiming when they hit the opponent (sometimes to trick their opponent as well). \textit{“I feel like the part of not knowing what other people had enhanced the experience because it made the game really funny. We both kept yelling random numbers and I don't think either of us were even close to what we said”} (P4). In addition, not being able to see the opponent's ball encouraged greater movement. \textit{"I had to guess which way the ball was going, so I kept moving since I figured the ball was aimed at where I was just standing"} (P5). Hiding details of the opponent from participants created a need for participants to communicate amongst themselves and encouraged more movement.

\subsection{The competitive nature of the game induced physical activity among participants and encouraged future play}
While there is no timer in the game, participants felt an implicit time due to the need to score before their opponent to win. As a result, many participants noted that the deeming of a winner and a loser at the end of a Ball-AR game led to greater movement and physical activity:  \textit{“I kept running after my ball to throw it again. Otherwise, I probably would have lost”} (P2). Other participants expressed their desire to play again after losing. 

\subsection{Participants felt that the blend of mobile AR and physical activity was fitting and memorable, though potentially dangerous }
Participants expressed that AR made the game more enjoyable and felt that it allowed for features normal mobile games could not. \textit{“I think AR should be geared towards actual real movements, not just sitting and playing with your finger. In the game, there is a physical element and it is like a sport”} (P6). In addition, the need for movement created a memorable experience within AR. \textit{"A lot of AR is memorable but this one had movement involved and that made it cooler than other AR simulations where you’re just looking at a virtual space"} (P10). However, some participants noted that the movement in the game and the small screen could potentially be dangerous. \textit{"Someone could definitely bump into tables and chairs and stuff if they didn't really know the room that well. Normally, it's easier not to, but since you're running and looking at a phone, I could see that happening"} (P2).

\subsection{Participants did not focus on the physical features of the environment and desired more augmentation} 
The ability to throw the ball off physical surfaces was included to ground the players in their physical environment and to foster belonging by connecting them through the shared space. However, most participants noted that they were more focused on other aspects of the game rather than the shared physical space. \textit{"I wasn't really that aware of my surroundings because I was mostly looking for my ball or looking at [P2] to hit [them]"} (P1). Participants also expressed that the main use of the feature was for strategy and scoring rather than feeling connected to a certain physical space with their opponent. In addition, participants actually requested greater augmentation to enhance the physical environment. \textit{“It would be cool if there were more cool things to throw off of in the room, like some virtual objects"} (P7).

\section{Discussion}
Through these results, we can theorize at good and bad design choices for AR games, specifically when fostering physical activity and co-located interaction. 

For physical activity, we discovered that giving each player one ball to follow around did induce movement. In addition, the competitive nature of the game fostered quicker movement due to the desire to win. In general, there seemed to be a sort of expectation among participants for AR games to expand on the capabilities of non-AR mobile games by encouraging movement. Most participants felt that the inclusion of physical movement and activity was fitting for AR and took advantage of AR's physical elements, hinting at the possibility that physical movement of some kind belongs in mobile AR games, although with more safety precautions.

In terms of co-located interaction, the use of hidden information fostered interesting communications among participants. In addition, most participants enjoyed being in the same physical space as their opponent as it allowed for these communications and created more enjoyable physical and social interactions. However, many participants had the belief that mobile games such as this should have some sort of single-player capability. This may be due to participants wanting to play on their own time without restriction or may be a result of the abundance of single-player mobile games currently available. 

We also noticed that physical activity and co-located interaction intersected greatly within the game. The encouragement of physical activity created more memorable experiences for users and fostered physical interaction. Participants had fun both running and seeing their opponent run frantically, and many recalled these moments afterwards as their favorite moments of the game. These results describe a potential correlation between physical activity in AR games and fun physical interaction between players that we encourage future mobile AR game designers to consider.

Lastly, many participants expressed the desire for more augmentation and greater immersion in the shared virtual space. This speaks to the potential preference of users to escape or augment reality while playing an AR game, rather than just interact with it. The features of the specific shared physical space did not make a substantial impact on the interaction or belonging of participants, and perhaps a greater virtual presence could create a more memorable experience for users. 

\section{Limitations}
While we determined both benefits and challenges in mobile AR game design, there were some limitations. Our findings come from a small, homogeneous group of participants, as all are college students with some interest in physical activity games. Future research should explore a broader group of participants. Furthermore, all studies were conducted in the same physical location. Future work should also experiment with different types of physical locations to expand this research. 

\section{Conclusion}
In this paper, we presented Ball-AR, an AR game designed to foster physical activity, co-located interaction, and physical environment utilization. We designed the game using both techniques from previous studies and new designs to encourage movement, increase communication, and ground users in their shared physical space. We then evaluated the game with five dyads. From our results, we learned that the combination of a shared physical space and physical movement created enjoyable interactions and that certain design features, such as competition, hidden information, and usage of AR were effective in increasing communication and activity. However, participants desired more augmentation and more usage of a virtual shared space.

Future works in designing co-located AR gaming experiences could expand on the results shown in this paper. For example, future work could study more specific types of physical activity or study the impacts of greater augmentation on co-located interaction. Overall, we hope future researchers continue to explore mobile AR games and their ability to positively impact interaction and health of players.
\bibliographystyle{ACM-Reference-Format}
\bibliography{ballar}

\appendix

\section{Interview Protocol}
Participants followed along the protocol shown below:
\begin{enumerate}
\item  Each of the 10 participants was asked to meet at a predetermined location, where they were told about the study and shown the instructions to the game.
    
\item Participants were separated into five groups of two. Each group played a best-of-three match (first to two wins) of Ball-AR.
\item After the match ended, each participant was individually interviewed. They were asked the following questions:
     \begin{itemize}
      \item What are your thoughts on the Ball-AR experience? Would you play again?
      \item What was your favorite feature of the game? What was your least favorite feature of the game?
      \item Did you enjoy the aspect of playing against another person? Would you have preferred a single-player game?
      \item Did you like the competitiveness of the game? Would you have preferred a collaborative game?
      \item Did playing the game in the same room as your opponent make the experience more enjoyable? Would you have preferred a virtual/remote version?
      \item Did you move at all while playing Ball-AR? How did physical movement affect your overall experience?
      \item Did you make use of the ability to bounce the ball off physical surfaces? What were your opinions on this feature? 
      \item Did the use of a mobile device make the game more enjoyable or less enjoyable? Why?
      \item Would you have preferred sharing a ball with your opponent rather than each having your own?
      \item Did you encounter any dangers while playing Ball-AR?
      \item Do you have any general suggestions for Ball-AR and its next steps?
    \end{itemize}
\end{enumerate}

\end{document}